\documentclass[showpacs,preprint,aps]{revtex4}
\usepackage{graphicx}

\begin{document}
\begin{center}
{\bf \Large  Momentum distribution of N$^*$ in nuclei}
\vskip 5mm
N. G. Kelkar\\
Departamento de Fisica, Universidad de los Andes, Cra 1E, 18A-10, 
Bogot\'a, Colombia\\
\end{center}
\begin{abstract}
Due to its dominance in the low energy 
eta-nucleon interaction, the S11 N$^*$(1535) resonance 
enters as an important ingredient in the analyses 
of experiments aimed at finding evidence for the existence of eta-mesic nuclei. 
The static properties of the resonance get modified inside the nucleus and 
its momentum distribution is used in deciding these properties as well as 
the kinematics in the analyses. Here we show that given the possibility for 
the existence of an N$^*$-$^3$He quasibound state, the relative momentum 
distribution of an N$^*$ and $^3$He inside such a 
$^4$He is narrower than that of neutron-$^3$He in 
$^4$He. Results for the N$^*$-$^{24}$Mg system are also presented. 
The present exploratory work could be useful in motivating searches of exotic 
N$^*$-nucleus quasibound states as well as in performing analyses of eta meson 
production data. 
\end{abstract}
\maketitle
\section{Introduction}
A few decades ago, a new topic in meson physics drew the attention of 
intermediate energy nuclear physicists. This was due to the finding that 
the interaction between the eta meson ($\eta$) and a nucleon is 
strongly attractive \cite{bhalerao} and 
that this interaction may generate sufficient attraction to give rise to
an exotic bound state (also referred to as ``quasibound" since it decays 
within a short time)  
when put in the
nuclear environment. The prediction for the existence of such eta-mesic 
nuclei initiated lots of efforts on the experimental as
well as the theoretical front \cite{otherreviews, ourreview}. 
Due to the lack of eta beams (as
the eta meson is extremely short lived),  
experiments where the $\eta$ was produced 
in the final state with protons and photons incident on nuclei, were 
performed. However, apart from two controversial experiments \cite{mainz}, 
there has been no definite evidence for the existence of these states. 
Meanwhile, the interest has also shifted from $\eta$ to $\eta^{\prime}$ 
mesic nuclei \cite{etaprime}. However, the 
WASA group \cite{wasapapers} is still active in the search for  
eta-mesic states in light nuclei (see also \cite{lightnucl} for 
theoretical works on eta-mesic helium nuclei).  

Many a time in physics, an experimental finding is not a direct measurement 
but rather a result deduced from the analysis of experimental data using 
theoretical inputs. For example, nuclear radii are not ``measured" 
but rather extracted \cite{mccarthy} using 
theoretical relations involving electromagnetic form factors of nuclei which 
are deduced from data on electron-nucleus scattering.
The experimental searches for eta-mesic nuclei involve certain assumptions 
and theoretical inputs too. One of the (sufficiently justified) assumption is 
that the interaction of the $\eta$ meson with the nucleus proceeds through 
the formation of the S11 N$^*$(1535) resonance.  
Hence, analyses of an anticipated eta mesic nucleus, model the eta-nucleon 
interaction to
proceed via the formation of an N*(1535) resonance which repeatedly decays,
regenerates and propagates within the nucleus until it eventually decays into a free meson and a nucleon. The search for an $^4$He-$\eta$ bound state which 
for example involves the analysis of the $d \,d \,\to \, ^3$He $\, N 
\pi$ reaction data, is performed by assuming that the reaction 
proceeds as follows \cite{magdathesis}: 
$d \,d \,\to (^4$He-$\eta)_{bound} \, \to \, (^3$He-N$^*) \,\to$
$^3$He $\, N \pi$. Thus it becomes necessary to incorporate the static 
properties and motion of the N$^*$ resonance inside the nucleus. 
One essential ingredient in these analyses is the relative momentum 
distribution of N$^*$-$^3$He inside the $^4$He nucleus 
(which contains an N$^*$ in place of one proton or neutron). 
This distribution is necessary to establish the 
detector system acceptance for the registration of the 
$d \,d  \, \to \, (^3$He-N$^*) \,\to$
$^3$He $\, N \pi$ reaction and to determine the data selection criteria  
\cite{wasapapers}. 
However, with the knowledge of the N$^*$ interaction with nucleons 
not being sufficient (see however the discussion in the 
next section), it is common to use the momentum distribution of a 
nucleon inside the nucleus rather than that of the resonance.
In fact, even though the momentum distributions inside nuclei provide 
information which is complementary to that obtained from electromagnetic 
form factors, much less experimental information is available on the 
former even in normal nuclei.  

In the present work, a model for the evaluation of the momentum distribution 
of an N$^*$ inside a nucleus is presented. In a recent work \cite{actaphysb}, 
the possibility 
for the existence of broad N$^*$-nucleus (quasi)bound states was proposed  
using some available sets of 
coupling constants for the N N$^*$ $\to$ N N$^*$ interaction. Since a 
few bound states 
in the N$^*$-$^3$He and N$^*$-$^{24}$Mg were indeed 
predicted, in this work we use these binding energies 
(as well as some others obtained by varying the coupling constants) to evaluate the 
momentum distribution of the N$^*$ resonance in these nuclei. 
In the next section, we shall briefly repeat the formalism used in 
\cite{actaphysb} and proceed further to describe the evaluation of the 
momentum distributions. An interesting outcome of these investigations 
is that the momentum distribution of an N$^*$ resonance inside a nucleus 
is narrower than that of a nucleon inside a nucleus. This fact could 
indeed be of significant importance in the analyses done in connection 
with the searches for eta-mesic nuclei.

\section{Model for the N$^*$-nucleus potential}
Though the existence of a bound state of a baryon resonance and a 
nucleus is by itself an exotic idea, it has indeed been explored 
in context with the $\Delta$ (spin-isopsin 3/2) resonance  
\cite{deltas} in the past. 
In \cite{dillig}, the author calculated the momentum distribution of 
such a resonance too. As compared to the $\Delta$, the case of the 
N$^*$(1535) resonance is relatively simpler. 
It is a spin 1/2 (negative parity) S11 resonance which decays dominantly 
into a nucleon and a pion or eta meson. Hence, 
we shall use a one meson exchange N N$^* \to$ N N$^*$ interaction with the 
exchange of a $\pi$ and $\eta$ meson. 
The N$^*$-nucleus potential is then obtained by folding the elementary 
N N$^*$ interaction with a nuclear density (see some remarks regarding the 
validity of the folding model in this work, above Eq.(\ref{potn})). 
We shall also retain the 
scalar part of the interaction only. 
Since the N$^*$(1535) is a negative parity baryon, indeed in the
one-pion and -eta exchange diagrams, the spin dependent terms are 
suppressed as compared to the leading scalar terms. 

As for the $\pi$NN$^*$ and $\eta$NN$^*$ coupling 
constants, there appears a range of values in literature \cite{couplings,
roebigaver, vetmoal, ansagh, kanchan, osetgar, carras}.
In the first reference in \cite{roebigaver}, for example, 
the cross sections for photoproduction of 
$\eta$ mesons from heavy nuclei were measured and compared with models 
of the quasifree $A(\gamma,p)X$ reaction. The authors adjusted the
coupling constants from an Effective Lagrangian Approach (ELA) in 
\cite{carras} to reproduce the $p(\gamma,\eta)p$ 
and $d(\gamma,\eta)np$ data. With $g_{\pi N N^*}$ = 0.699 and 
$g_{\eta N N^*}$ = 2.005, the experimental $\eta$ photoproduction cross sections
on complex nuclei were reproduced within the model of \cite{carras}. 
In the two references in \cite{vetmoal}, the authors found $g_{\pi N N^*}$ = 0.8 
and $g_{\eta N N^*}$ = 2.22 while comparing the calculations within a one boson exchange 
model with the $ N N \to N N \eta$ and $\pi^- p \to \eta n$ data. 
Somewhat bigger values of the ${\pi N N^*}$ coupling constant have been found in more 
recent years with Ref. \cite{ansagh} for example, reporting $g_{\pi N N^*}$ = 1.09 
by comparing calculations within a chiral constituent quark model with the experimental 
data on the partial decay width of the S11(1535) resonance. Mixing pseudoscalar 
meson-baryon with vector meson-baryon states in a coupled channels scheme with 
$\pi N$, $\eta N$, $K \Lambda$, $K \Sigma$, $\rho N$ and $\pi \Delta$, the 
coupling constants, $g_{\pi N N^*}$ = 1.05 and $g_{\eta N N^*}$ = 1.6 were obtained 
in \cite{osetgar}. In a study of nonstrange meson baryon systems 
where the N*(1535) was found to get  generated as a result of coupled channel 
dynamics of vector meson-baryon and pseudoscalar-baryon systems, 
the authors \cite{kanchan} obtain, $g_{\pi N N^*}$ = 0.95 and 
$g_{\eta N N^*}$ = 1.77. We shall present results with some sets of 
coupling constants mentioned above. The constants and binding energies of 
possible N$^*$-$^3$He states are listed in Table I. 
As compared to the $\pi$N N$^*$ and $\eta$N N$^*$ couplings, the 
$\pi$N$^*$N$^*$ and $\eta$N$^*$N$^*$ couplings 
are even much less known. In view of the above uncertainties and also the fact that 
the present work is aimed at finding out how much the N$^*$ momentum 
distribution in a nucleus differs from that of a nucleon, we do not attempt
a more sophisticated calculation. 

\subsection{Elementary N N$^*$ interaction}
The elementary interaction is considered to proceed by the exchange of a pion
and an eta meson as shown in  Fig. 1. We consider an N$^*$
which is neutral. 
The calculation for a positively charged N$^*$ can be repeated in
a similar way.
\begin{figure}[h]
\begin{center}
\includegraphics[width=12cm,height=5cm]{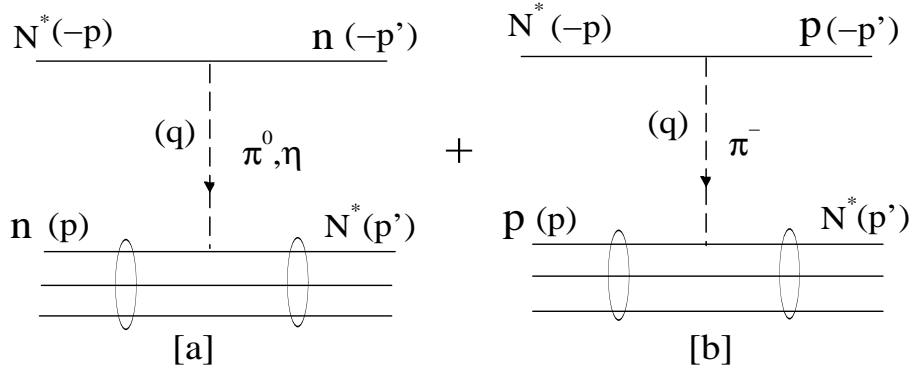}
\caption{\label{fig:eps1} Elementary N N$^*$ $\to$ N N$^*$ processes
considered in the interaction of the N$^*$ with a nucleus.}
\end{center}
\end{figure}
Diagrams involving 
the N$^*$N$^*\,\pi$ or N$^*$N$^* \,\eta$ couplings which are hardly known 
will not be considered. 
Apart from this fact, for such diagrams, the potential turns out to
be spin dependent (and so also suppressed as compared to the leading 
term in the potential of Fig. 1 ).

The $\pi$NN$^*$ and $\eta$NN$^*$ couplings 
(with N$^*$(1535,1/2$^-$)) are given by
the following interaction Hamiltonians \cite{osetetaNN}:
\begin{eqnarray}\label{hamil}
\delta H_{\pi N N^*} = g_{\pi N N^*} \bar{\Psi}_{N^*}  {\vec \tau} \Psi_N \cdot 
{\vec \Phi_{\pi}} + {\rm h.c.}\\ \nonumber
\delta H_{\eta N N^*} = g_{\eta N N^*} \bar{\Psi}_{N^*}  \Psi_N \cdot 
\Phi_{\eta} + {\rm h.c.}
\end{eqnarray}
Let us consider the diagram for the N$^*$ n $\to$ n N$^*$
process in Fig. 1 and use the standard Feynman diagram rules with the
non-relativistic approximation for the spinors
\begin{equation}\label{spinoreq3}
u_i =\sqrt{2m_i}\left(\begin{array}{c} w_i\\
{\vec{\sigma}_i \cdot \vec{p}_i \over 2m_ic}\, w_i 
\end{array}\right) \, ,
\end{equation}
to write the amplitude as
\begin{equation}
{g_{xNN^*}^2 \bar{u}_{N^*}(\vec{p}^{\, \prime}) \, u_n(\vec{p}) \,
\bar{u}_n(-\vec{p}^{\, \prime})\, u_{N^*}(-\vec{p}) \over q^2 - m_x^2}\, ,
\end{equation}
where $x = \pi$ or $\eta$ and $q^2= \omega^2 - \vec{q}^2$ is the four momentum 
squared carried by the exchanged meson ($q = p^{\prime} - p$ as shown in the figure). 
Here for example,
\begin{equation} 
\bar{u}_n(-\vec{p}^{\, \prime})\, u_{N^*}(-\vec{p}) = N \, \biggl( 1 \, -\, 
{\vec{\sigma}_n \cdot \vec{p}^{\,\prime} \vec{\sigma}_{N^*} \cdot \vec{p} \over 
4 m_N m_N^* c^2} \biggr ) 
\end{equation}
and we drop the second term in the brackets which is spin dependent as well as
$1/c^2$ suppressed.
The potential in momentum space obtained from the above amplitude is given as:
\begin{equation}\label{pot1}
v_x(q) = {g^2_{xNN^*} \over q^2 - m_x^2} \, 
\biggl ({\Lambda^2_x - m_x^2 \over \Lambda_x^2 - q^2} \biggr )^2\, , 
\end{equation}
where the last term in brackets has been introduced to take into account the
off-shellness of the exchanged meson. The four momentum transfer squared,  
$q^2 = \omega^2 - \vec{q}^2$, in the present calculation is approximated 
simply as 
$q^2 \simeq - \vec{q}^2$. Since the mass of the N$^*$ is much bigger than that 
of the nucleon, the neglect of the energy transfer, $\omega$, in the 
elastic N N$^*\, \to$ N N$^*$ process as such is not well justified. 
However we do not expect the relative momentum distribution of the 
N$^*$ in the nucleus to depend strongly on the mass of the N$^*$ 
(an expectation which will be verified later numerically). We thus proceed 
further without a non-zero $\omega$  which would give rise to poles in 
(\ref{pot1}) and make the calculation of the N$^*$ nucleus 
potential a formidable task. 
The potential in (\ref{pot1}) is Fourier transformed to obtain the 
potential in $r$-space. 
The Fourier transform of (\ref{pot1}) can be calculated analytically and we get,
\begin{equation}\label{potelement}
v_x(r) = {g^2_{xNN^*} \over 4 \pi} \,\biggl [ {1\over r} \biggl 
( e^{-\Lambda_x r} - e^{-m_x r} \biggr ) + {\Lambda_x^2 - m_x^2 \over 2 \Lambda_x} \, 
e^{-\Lambda_x r} \biggr ]\, .
\end{equation}
In order to evaluate the above potential, we need to know the coupling 
constants at the $\pi$NN$^*$ and $\eta$NN$^*$ vertices. One can 
find a range of values in literature as discussed above. 
In Fig. 2, we see the sensitivity of these potentials to the use of different 
sets of parameters. Whereas the first two sets are shown to display the 
sensitivity to the values of the cut-off parameters, the next set is the one 
which gives the highest binding of the N$^*$ and nuclei in this work. 
It gives 
rise to one bound N$^*$-$^3$He state at -4.78 MeV and 3 bound N$^*$-$^{24}$Mg
states at -50.3, -22.5 and -3.25 MeV. Using this 
as well as other sets listed in Table I, we perform an 
exploratory study of the momentum distributions of the N$^*$.  

\begin{figure}[h]
\begin{center}
\includegraphics[width=9cm,height=11cm]{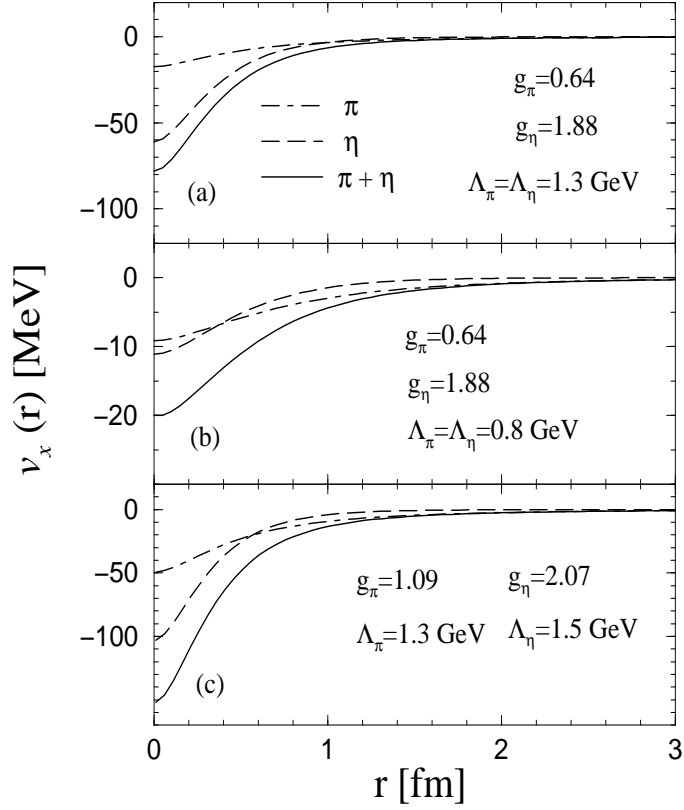}
\caption{\label{fig:eps2} Elementary potential as given in 
Eq.(\ref{potelement}). Note that whereas $\pi + \eta$ exchange contributes to 
the n N$^* \to$ n N$^*$ potential, only $\pi$ contributes to the 
p N$^* \to$ p N$^*$ potential in N$^*$-$^3$He.}
\end{center}
\end{figure}

\subsection{N$^*$-nucleus potentials}
Once the elementary potential has been defined, the folding model with 
\begin{equation}
V(R) = \int \, d^3r\, \rho(r) \, v(|\vec{r} - \vec{R}|) \, , 
\end{equation}
is used to construct the N$^*$ nucleus potential $V(R)$ which is given by 
\begin{eqnarray}\label{nuclpot}
V(R) &=& V_p(R) + V_n(R) \nonumber \\ 
&=& Z \,\int \, d^3r\, \rho_p(r) \, v_p(|\vec{r} - \vec{R}|) \, +\, 
N \,\int \, d^3r\, \rho_n(r) \, v_n(|\vec{r} - \vec{R}|) \,, 
\end{eqnarray}
where, $Z$ and $N$ are the number of protons and neutrons,
$v_n(r) = v_{\pi^0}(r) + v_{\eta}(r)$ and
due to the isospin factor appearing in the $\pi^-$ exchange diagram
(see Fig. 1 and Eq.(\ref{hamil})),
$v_p(r) =  v_{\pi^-}(r) \vec{\tau}_1 \cdot \vec{\tau}_2$.
Note that in the case of $^3$He with $Z = 2$ and due to the isospin factor in 
$v_p(r)$, the contribution of $V_p(R)$ to the total $V(R)$ is much larger 
than that of $V_n(R)$. Since $v_p(r)$ (and hence $V_p(R)$) involves only the 
pion exchange diagram (Fig. 1[b]), the dominant contribution to 
the N$^*$-nucleus potential comes from pion exchange. Since the pion exchange 
potential is fairly long range, the folding model chosen in the present work, 
though not ideal, seems acceptable. 

After performing the angle integration, the above integral reduces
for example to
\begin{eqnarray}\label{potn}
V_n(R) = {-2 \pi A \over R} \, \int \, \biggl \{ 
{e^{-m_x (|r - R|)} - e^{-m_x (r + R)} \over m_x} \, - \, 
{e^{-\Lambda_x (|r - R|)} - e^{-\Lambda_x (r + R)} \over \Lambda_x} 
\nonumber \\
+ B \biggl [ \, \biggl({r+R \over \Lambda_x} +{1 \over \Lambda_x^2} \biggr ) 
\,e^{-\Lambda_x(r+R)}\, -\, \biggl ( {|r-R| \over \Lambda_x} + {1 \over 
\Lambda_x^2} \biggr ) \,e^ {-\Lambda_x |r - R|}\, \biggr]  
\,\biggr \}\,
r \, dr\, \rho_n(r), 
\end{eqnarray}
where $A = g^2_{xNN^*}/4\pi$ and $B= (\Lambda_x^2 - m_x^2)/2\Lambda_x$.

In case of the $^3$He nucleus, the majority of information available in 
literature is on the charge density distribution of $^3$He obtained from 
electron scattering. The root mean square radius, $r_{ch}^{3He}$ = 1.88 $\pm$ 0.05 fm, 
obtained in \cite{mccarthy} from the $^{3}$He charge form factor 
is a bit smaller than the value of 1.959 $\pm$ 0.03 fm in \cite{amroun}. Ref. 
\cite{amroun} also provides the charge form factor of $^3$H with 
$r_{ch}^{3H}$ = 1.755 $\pm$ 0.086 fm. 
There exists a parametrization of the matter density given in \cite{cookgrif} where 
a folding model analysis of $^3$He elastic scattering on heavy nuclei is performed. 
The authors fit the parameters in a gaussian density to reproduce a 
$^3$He matter radius of 1.68 fm (calculated as $r_{mat}^2 = r_{ch}^{3He} - r_p^2$ 
with $r_p$ being the radius of the proton). 
There is however no direct experimental data for the neutron density distribution 
in $^3$He. We identify the neutron density distribution in $^3$He with the 
proton density in $^3$H, which not only seems reasonable provided that the charge
symmetry breaking is small but also agrees with the matter distribution given in 
\cite{cookgrif}. Such an approach of calculating the nuclear densities using the 
charge densities of $^3$He and $^3$H has also been used earlier 
in literature \cite{tsushima}. 
Thus, for the proton density distribution $\rho_p(r)$, we choose 
a sum of Gaussians \cite{amroun}, namely,  
\begin{equation}\label{helidensity}
\rho(r) = {1 \over 2 \pi^{3/2} \gamma^3} \sum_{i=1}^N \, 
{Q_i \over 1 + 2R_i^2/\gamma^2} \, \biggl ( e^{-(r-R_i)^2/\gamma^2} + 
e^{-(r+R_i)^2/\gamma^2} \biggr)\, , 
\end{equation}
where the parameters $Q_i$, $R_i$ and $\gamma$ for 
$^3$He and $^3$H can be found in \cite{amroun}. 
Thus, with $\rho_p = \rho_{ch}^{3He}$ and $\rho_n = \rho_{ch}^{3H}$ 
(both normalized to 1), the above integral can in principle be done analytically. 
However, the analytic results are lengthy expressions which include  
error functions and exponentials. They are not particularly
enlightening and hence we rather perform the integral numerically.
The density for $^{24}$Mg
is assumed to have the following Woods-Saxon form \cite{magdensity}:
\begin{equation}\label{woodensity}
\rho(r) = {\rho_0 \over 1 + \exp{\biggl({r - c\over a} \biggr)}}\, ,
\end{equation}
where $c = r_A\, [1 - (\pi^2 a^2/ 3 r_A^2)]$ with $a = 0.54$ fm and
$r_A = 1.13 A^ {1/3}$.
The N$^*$ nuclear potentials thus evaluated (see \cite{actaphysb}) 
can be fitted reasonably well to Woods Saxon
forms of potentials. This fact facilitates the search for a possible 
N$^*$-nucleus bound state and the calculation of its wave function and 
hence momentum distribution. The potentials corresponding to the 
various sets of parameters in Table I can be fitted by a Woods Saxon potential 
with the depth parameter $V_0$ 
ranging between 14 to 42 MeV, $a$ = 0.8 fm and $R$ from 1.15 to 1.34 fm.

\begin{table}[ht]
\caption{ The $\pi N N^*$ and $\eta N N^*$ coupling constants and the
binding energies of the possible N$^*$-$^3$He bound states obtained with 
the corresponding set in the N N$^*$ $\to$ N N$^*$ potentials.}
\begin{tabular}{|l|l|l|l|}
  \hline
   & $g_{\pi N N^*}$ & $g_{\eta N N^*}$  & \, E (MeV)  \\
   &  &  &  \\
  \hline
Chiral constituent quark model   & \, 1.09 & \, 2.07  & \, -4.78 \\
fits partial decay widths \cite{ansagh}   &  &  &  \\
  \hline
  Hidden gauge formalism $^{\dagger}$ & \, 1.05 & \, 1.6  &\, -3.6  \\
fits partial widths and $\pi^- p \to \eta n$   \cite{osetgar} &  &  &  \\
  \hline
vector- and pseudoscalar-baryon    & \, 0.95  & \, 1.77  & \, -2.1 \\
coupled channel study \cite{kanchan} & & & \\
$^{\dagger}$ N$^*$(1535) is dynamically generated & & & \\
\hline
 One boson exchange model    & \, 0.8 & \, 2.22 & \, -0.8 \\
fits $ p p \to p p \eta$ data \cite{vetmoal}  & & & \\
\hline
 Data on $\eta$ photoproduction on heavy nuclei \cite{roebigaver}  
& \, 0.669 & \, 2.005  & \, -0.04  \\
fits $p(\gamma,\eta)p$
and $d(\gamma,\eta)np$ data  within ELA \cite{carras}&  &  &  \\
  \hline
\end{tabular}
\end{table}

\section{Momentum distribution of the N$^*$ in nuclei}
The Schr\"odinger equation for the Woods Saxon potential can be reduced to
one for the hypergeometric functions \cite{WShypergm} and a condition for
the existence of bound states can be found.
For a Woods Saxon potential of the type
\begin{equation}
V(r) = - {V_0 \over 1 + e^{r-R\over a}}
\end{equation}
the Schr\"odinger equation
\begin{equation}\label{schrodeq} 
{d^2u \over dr^2} + {2 \over r} {du\over dr} + {2m\over \hbar^2} (E - V) u =0  
\end{equation}
may be transformed to the independent variable
$y = 1 / [1 + e^{r - R/ a}]$ to obtain a hypergeometric differential equation. 
After some algebra \cite{WShypergm} one obtains the following
condition for bound states:
\begin{equation}\label{boundcond}
{\lambda R \over a} \, +\, \Psi \,-\, 2 \phi \, - \arctan{\lambda \over \beta} 
\, =\, (2n - 1) {\pi \over 2}\, \, \, \, \, n = 0, \pm 1, \pm 2, ...
\end{equation}
where,
$${2 m E \over \hbar^2} \, a^2 = - \beta^2 ; \,\,\, {2 m V_0 \over \hbar^2} \, 
a^2 =  \gamma^2 ; \, \, \, \lambda = \sqrt{\gamma^2 - \beta^2}$$
and $\phi = arg \Gamma (\beta + i \lambda)$; $\Psi = arg \Gamma (2 i \lambda)$.

Defining $u(r) = \chi(r)/r$ and $y = 1 / [1 + e^{r - R/ a}]$, the solution of 
the hypergeometric differential equation can be found to be
\begin{equation}
\chi = y^{\nu} \, (1 - y)^\mu \, _2F_1(\mu+\nu, \mu+\nu+1, 2\nu+1;y)
\end{equation}
where $\nu = \beta$ and $\mu^2 = \beta^2 - \gamma^2$. 
Since the variable $y$ is given in terms of $r$, we essentially have the 
wave function 
$\chi(r)$ which can then be Fourier transformed as follows:
\begin{equation}
\chi(p) = \biggl ( {2\over \pi} \biggr )^{1/2} \, \int_0^{\infty} \, r j_0(pr) \chi(r) 
dr 
\end{equation}
to evaluate the momentum distribution $T(p)$ as, 
\begin{equation}
T(p) = {1 \over 4 \pi} \, |\chi(p)|^2\,\,p^2 .
\end{equation}
$T(p)$ is normalized such, that, 
\begin{equation}
4 \pi \, \int \, T(p) \, dp = 1
\end{equation}
\begin{figure}[h]
\begin{center}
\includegraphics[width=9cm,height=9cm]{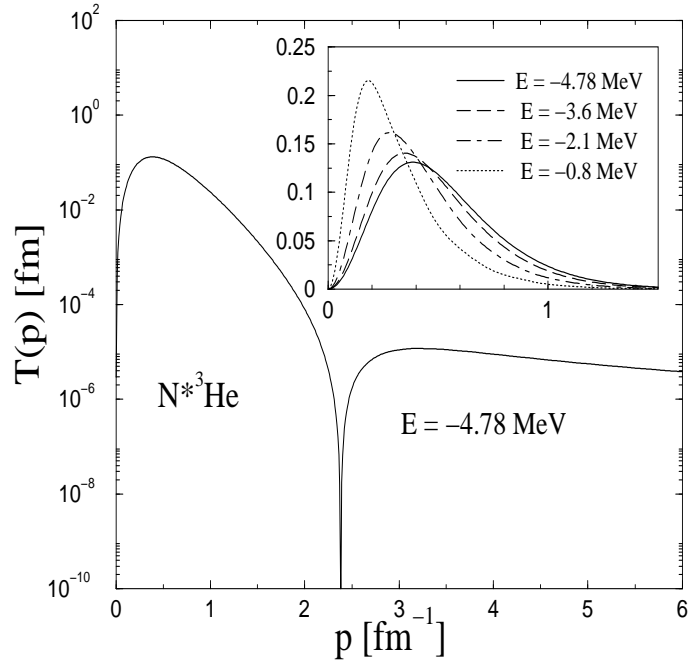}
\caption{\label{fig:eps3} Momentum distribution of the N$^*$ in $^4$He. 
Calculations with different sets of coupling constants (given in Table I) are
shown in the inset on a linear scale.}
\end{center}
\end{figure}
Figure 3 displays the relative momentum distribution $T(p)$ of N$^*$-$^3$He 
inside a $^4$He nucleus which contains an N$^*$ instead of a neutron. 
The curve shown on the logarithmic scale 
corresponds to the set of parameters in Table I which give the highest binding. 
The results for other parameter sets in Table I are shown on a linear scale 
in the inset since one does not see much difference at small momenta on the 
log scale. The offshell cut-off parameters appearing in the elementary N N$^*$ $\to$ 
N N$^*$ potential are chosen to be 
$\Lambda_{\pi}$ = 1.3 GeV and $\Lambda_{\eta}$ = 1.5 GeV in all cases. 
Changing the cut-offs to $\Lambda_{\pi}$ = 0.9 GeV and $\Lambda_{\eta}$ = 1.3 GeV
for example, does not change the distribution significantly (except for 
a small shift at high momenta) and is hence not shown in the figure. 

\subsection{Dependence on the N$^*$ mass}
The N$^*$-$^3$He potentials do not depend on the mass of the N$^*$ but in the 
search for bound states using the condition (\ref{boundcond}), 
one has to introduce the N$^*$ mass to calculate the reduced mass in 
that expression. 
In order to check the sensitivity of the results to the choice of the 
N$^*$ mass, 
we varied it between 1400 and 1550 MeV. The corresponding binding energies
of states fulfilling the condition (\ref{boundcond})
varied from 4.34 to 4.84 MeV for the parameter set chosen \cite{ansagh}. This 
variation introduces a very small change in the form of the bound 
state wave function as well as momentum distribution as can be seen 
in Fig. 4[a]. 
\begin{figure}[h]
\begin{center}
\includegraphics[width=16cm,height=8cm]{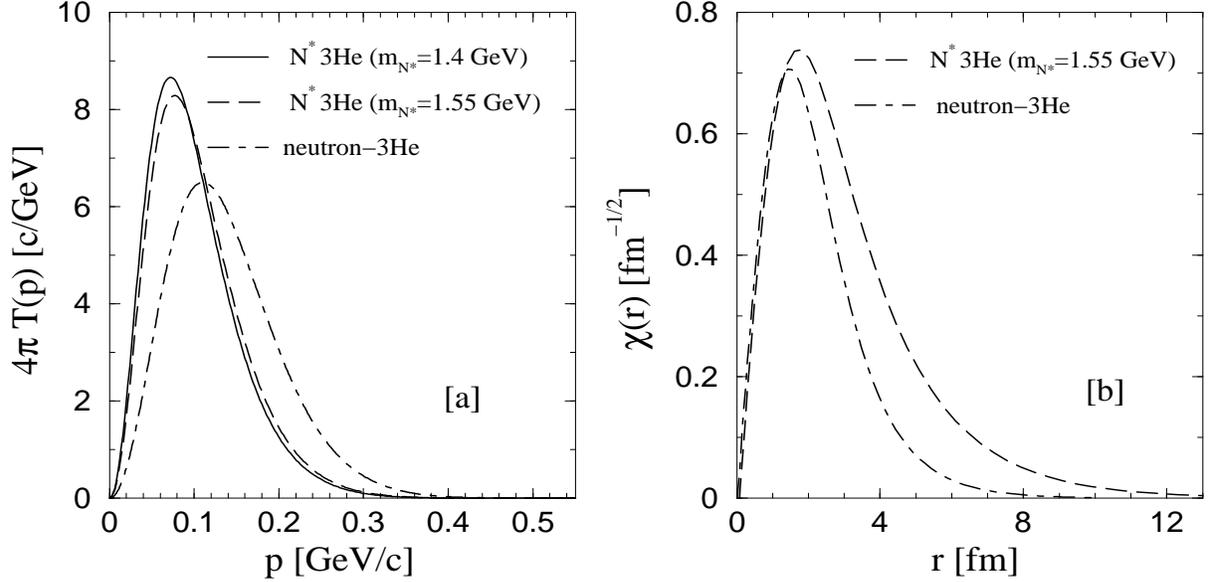}
\caption{\label{fig:eps4} Variation of the N$^*$-$^3$He momentum distribution 
in $^4$He with N$^*$ mass (solid and dashed lines correspond to 1400 and 
1550 MeV respectively in [a]). The dot-dashed 
line corresponds to [a] the momentum 
distribution and [b] the wave function of the neutron-$^3$He bound state 
(calculated within the same model). 
The wave function of N$^*$-$^3$He (dashed line) for m$_{N*}$ = 1550 MeV 
is also shown in [b].}
\end{center}
\end{figure}
This finding complements earlier results from \cite{roebigaver} which indicate 
little modification of the in-medium excitation of the S11(1535). 
Though some evidence of broadening was reported in \cite{yorita}, the N$^*$ mass 
of ~1544 MeV calculated in the Quark Meson Coupling (QMC) model (with the N$^*$ 
interpreted as a 3-quark state) \cite{bassthomas} seems to be consistent with 
the former experimental findings as well as the results of the present work.

\subsection{Comparison with a nucleon momentum distribution in $^4$He}
In order to compare the N$^*$-$^3$He relative 
momentum distribution in $^4$He with that of a nucleon in standard $^4$He, 
we replace the Woods Saxon parameters by $V_0$ = 66 MeV, 
$R$ = 1.97 fm and $a$ = 0.65 fm,  
to get a neutron-$^3$He potential which produces a state at -20.6 MeV
while fulfilling the condition in (\ref{boundcond}) with the reduced mass of 
a neutron and $^3$He. This is indeed close to the energy required to 
separate a neutron from $^4$He. Even if the curve for the momentum 
distribution of the neutron calculated in this manner does not have the 
authenticity of one evaluated using few body equations, it is pretty close
to a realistic calculation \cite{nogga} (see Fig. 5 and the discussion below) 
and serves for the purpose of 
comparison. In Fig. 4[b] we see the difference between the 
bound wave functions for the N$^*$-$^3$He 
and neutron-$^3$He systems which explains the difference in the distributions 
in Fig. 4[a]. With the N$^*$-$^3$He being loosely bound (-4.78 MeV) (as 
compared to the neutron which is bound by -20.6 MeV), the wave function 
of the N$^*$-$^3$He is more spread out in $r$-space (Fig. 4[b]). This causes 
the momentum distribution to be narrower. Other sets of 
parameters for the N N$^*$ interaction leading to lesser binding lead to even  
narrower distributions as seen in the inset in Fig. 3. 
A better agreement on the 
$\pi$NN$^*$ and $\eta$NN$^*$ coupling constants would be useful in order 
to perform a more accurate estimate of the momentum distribution of the N$^*$ in the 
nucleus.
\begin{figure}[ht]
\begin{center}
\includegraphics[width=9cm,height=9cm]{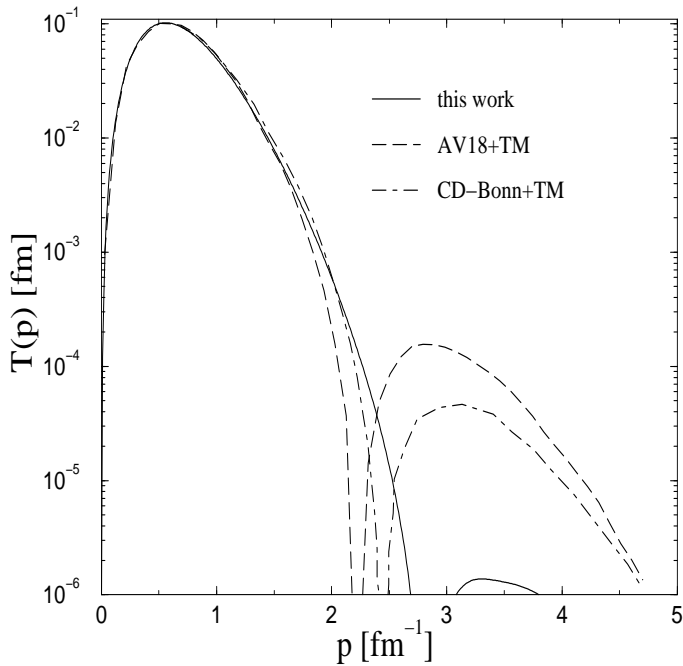}
\caption{\label{fig:eps5} Comparison of the proton-$^3$H momentum distribution 
in $^4$He of the present work (solid line) with 
others calculated using Faddeev-Yakubovsky equations \cite{nogga} 
with the AV18+TM (dashed) and CD-Bonn+TM (dot-dashed) potentials.}
\end{center}
\end{figure}

In order to test the validity of the calculations done in the present work, 
we repeat a similar calculation for the proton-$^3$H system 
in $^4$He for which some results using few body equations exist in literature. 
Though the momentum distribution for $n$-$^3$He is not expected 
to be very different from that of $p$-$^3$H in $^4$He, we perform this calculation
in order to compare with the available few-body results. 
With the Woods Saxon parameters of $V_0$ = 66 MeV,
$R$ = 1.93 fm and $a$ = 0.65 fm which reproduce the $p$-$^3$H binding of 19.8 MeV, 
we obtain a distribution which agrees at small and medium momenta with more 
sophisticated calculations \cite{nogga} shown in Fig. 5. The disagreement is 
only in the region of large momenta where the magnitude of $T(p)$ has fallen 
down by three orders of magnitude. Thus, the conclusion that the 
N$^*$-$^3$He momentum distribution is narrower than the neutron-$^3$He distribution, 
drawn from the calculations of the present work seems quite reliable.

\begin{figure}[ht]
\begin{center}
\includegraphics[width=9cm,height=9cm]{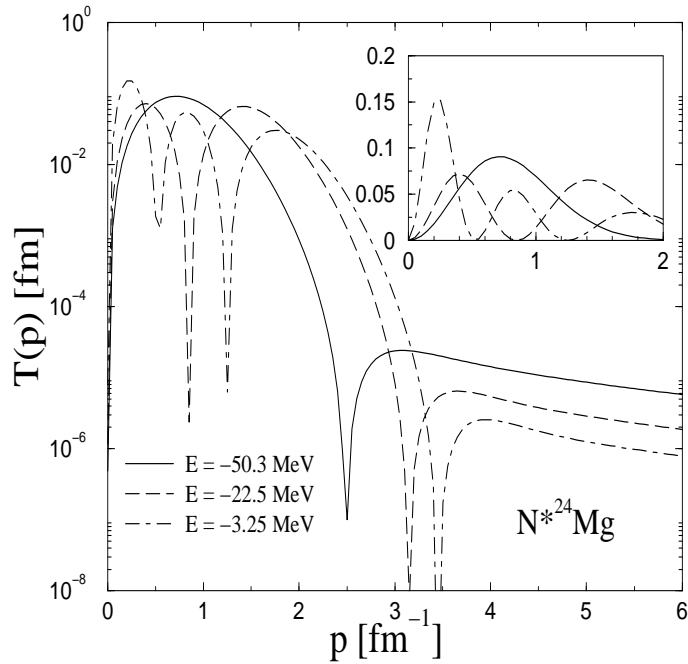}
\caption{\label{fig:eps6} N$^*$-$^{24}$Mg momentum distribution 
in $^{25}$Mg for three possible binding energies corresponding 
to the number of nodes $n = 1$, $2$ and $3$ in the bound wave functions.}
\end{center}
\end{figure}

\subsection{N$^*$-$^{24}$Mg bound states}
Using the set of parameters from \cite{ansagh} for the $\pi$NN$^*$ and 
$\eta$NN$^*$ coupling constants, the condition (\ref{boundcond}) 
allows three bound states of N$^*$-$^{24}$Mg 
at energies of -50.3, -22.5 and -3.25 MeV for 
n = 1, 2 and 3 respectively. The 
Woods Saxon potential parameters of the N$^*$-$^{24}$Mg system are: 
$V_0$ = 80 MeV, 
$a$ = 0.97 fm and $R$ = 2.85 fm.  In Fig. 6 one can see the distributions with 
one, two and three nodes accordingly.

\section{Summary}
The broad S11 baryon resonance N$^*$(1535) enters as one of the most essential 
ingredient in reactions involving the production of the 
neutral pseudoscalar eta meson ($\eta$) 
and hence also in the analyses of possible 
eta-mesic nuclei. Since the low energy $\eta$N interaction predominantly 
proceeds by producing an N$^*$ resonance which propagates, decays and 
regenerates inside the nucleus, it seems legitimate to ponder about 
the possible existence of an N$^*$-nucleus bound state too. Indeed, 
performing such an investigation in \cite{actaphysb}, it was found that
depending on the strength of the N N$^*$ interaction, loosely bound, 
broad quasibound states of the N$^*$ with $^3$He and $^{24}$Mg nuclei 
can be formed. In the present work, the investigation is continued to 
evaluate the momentum distribution of such an N$^*$ inside the nucleus.
Being aware of the fact that neither does any 
experimental evidence 
of N$^*$-nuclei exist 
nor is the N N$^*$ interaction accurately known, the calculations 
are done within a 
folding model where the elementary N N$^*$ $\to$ N N$^*$ 
potential is folded with the known nuclear densities. 
The present work finds that since the N$^*$ is loosely 
(or even very loosely, 
depending on the $\pi$NN$^*$ and $\eta$NN$^*$ couplings) 
bound to a nucleus, the bound state 
wave function of an N$^*$ as compared to that of a nucleon is more spread 
out in $r$-space and hence the momentum distribution is narrower than 
in case of the nucleon.
This finding is important in view of the fact that experimental analyses 
generally approximate the momentum distribution of an N$^*$ by that of a 
nucleon in a nucleus. The present work is a first attempt to 
evaluate the N$^*$(1535) resonance momentum distribution in nuclei. 
This distribution, as mentioned in the beginning, is necessary to establish the
detector system acceptance for the registration of the
$d \,d  \, \to \, (^3$He-N$^*) \,\to$
$^3$He $\, N \pi$ reaction and to determine the data selection criteria
\cite{wasapapers}.
A calculation of the momentum distribution of N$^*$-d in $^3$He would be necessary 
for the analysis of the $p \, d  \, \to \, (d$-N$^*) \,\to$
d $\, N \pi$ reaction aimed at searching $\eta$-mesic $^3$He whose prospects 
seem higher due to the fact that we have already seen a strong enhancement 
in the $p \, d\,\to \, ^3$He $\eta$ reaction near threshold. 
Such a calculation would however be better performed using a few body formalism 
for the N$^*$-p-n system. 
An improved knowledge of the N$^*$ coupling constants and experimental 
searches of N$^*$ nuclei could motivate such sophisticated 
few body calculations in future.   

\begin{acknowledgments}
The author thanks Prof. Pawel Moskal for useful comments and discussions.  
The author also thanks the Faculty of Science at the University of 
Los Andes, Colombia for financial support 
(project no. P15.160322.009/01-01-FISI02).  
\end{acknowledgments}


\begin{thebibliography}{99}
\bibitem{bhalerao}
R. S. Bhalerao and L. C. Liu, Phys. Rev. Lett. 54, 865 (1985).
\bibitem{otherreviews}
H. Machner, {\it J.Phys. G} {\bf 42} 043001 (2015);
B. Krusche, C. Wilkin, {\it Prog. Part. Nucl. Phys.} {\bf 80}, 43 (2014);
Q. Haider and L. C. Liu, Int. J.Mod. Phys E {\bf 24}, 1530009 (2015). 
\bibitem{ourreview}
N. G. Kelkar, K. P. Khemchandani, N. J. Upadhyay and B. K. Jain,
{\it Rep. Prog. Phys.} {\bf 76}, 066301 (2013).
\bibitem{mainz}
M. Pfeiffer {\it et al}., Phys. Rev. Lett. {\bf 92}, 252001 (2004);
G. A. Sokol and L. N. Pavlyuchenko, Phys. Atom. Nucl {\bf 71}, 509 (2008).
\bibitem{etaprime} 
S. D. Bass, Hyperfine Int. {\bf 234}, 41 (2015); S.D. Bass and P. Moskal, 
Acta Phys. Pol. B {\bf 47}, 373 (2016); H. Fujioka {\it et al.} (for 
Super-FRS collaboration), Hyperfine Int. {\bf 234}, 33 (2015).  
\bibitem{wasapapers}
W. Krzemien, P. Moskal and M. Skurzok, {\it Acta Phys. Polon. B} 
{\bf 46} 757 (2015); M.Skurzok, W. Krzemien, O. Rundel and P. Moskal, 
EPJ Web Conf. {\bf 117}, 02005 (2016); P. Adlarson {\it et al}., Phys. Rev. C 
{\bf 87}, 035204 (2013). 
\bibitem{lightnucl}
S. Wycech, A. M. Green and J. A. Niskanen, Phys. Rev. C {\bf 52}, 544 
(1995); S. Wycech and A. M. Green, Int. J. Mod. Phys. {\bf A} 20, 637 
(2005); N. G. Kelkar, K. P. Khemchandani and B. K. Jain, J. Phys. G {\bf 32}, 
1157 (2006); N. G. Kelkar, Phys. Rev. Lett. 99, 210403 (2007).
\bibitem{mccarthy}
J. S. McCarthy, I. Sick and R. R. Whitney, Phys. Rev. C {\bf 15}, 1396 (1977).
\bibitem{magdathesis}
M. Skurzok, Ph.D. thesis, Jagiellonian University (2015), arXiv:1509.01385.
\bibitem{actaphysb}
N. G. Kelkar, D. Bedoya Fierro and P. Moskal, Acta Phys. Pol. B {\bf 47}, 
299 (2016). 
\bibitem{deltas}
P. Bartsch {\it et al}., Eur. Phys. J. A {\bf 4}, 209 (1999); 
T. Walcher, Phys. Rev. C {\bf 63}, 064605 (2001); 
C. Chumillas, A. Parre\~no and A. Ramos, Nucl. Phys. A {\bf 791}, 329 (2007).
\bibitem{dillig}
M. Dillig, Phys. Rev. C {\bf 14}, 2226 (1976). 
\bibitem{couplings}
A. B. Santra and B. K. Jain, Nucl. Phys. A {\bf 634}, 309 (1998); 
W. Peters, U. Mosel and A. Engel, Z. Phys. A {\bf 353}, 333 (1996);
Ju-Jun Xie, Bing-Song Zou and Huon-Ching Chiang, Phys. Rev. C {\bf 77},
015206 (2008); A. Fix and H. Arenh\"ovel, Nucl. Phys. A {\bf 697}, 277 (2002).
\bibitem{roebigaver}
M. R\"obig-Landau {\it et al}., Phys. Lett. B {\bf 373}, 45 (1996); 
R. Averbeck {\it et al}., Z. Phys. A {\bf 359}, 65 (1997).
\bibitem{vetmoal}
T. Vetter, A. Engel, T. Bir\'o and U. Mosel, Phys. Lett. B {\bf 263}, 153 (1991); 
A. Moalem, E. Gedalin, L. Razdolskaja and Z. Shorer, Nucl. Phys. A {\bf 589}, 
649 (1995).
\bibitem{ansagh}
C. S. An and B. Saghai, Phys. Rev. C {\bf 84}, 045204 (2011).
\bibitem{kanchan}
K. P. Khemchandani, A. Mart\'inez Torres, H. Nagahiro and A. Hosaka, 
Phys. Rev. D {\bf 88}, 114016 (2013).
\bibitem{osetgar}
E. J. Garzon and E. Oset, Phys. Rev. C {\bf 91}, 025201 (2015). 
\bibitem{carras}
R. C. Carrasco, Phys. Rev. C {\bf 48}, 2333 (1993).
\bibitem{osetetaNN}
B. Lopez Alvaredo and E. Oset, Phys. Lett. B {\bf 324}, 125 (1994).
\bibitem{amroun}
A. Amroun {\it et al}., Nucl. Phys. A {\bf 579}, 596 (1994). 
\bibitem{cookgrif}
J. Cook and R. J. Griffiths, Nucl. Phys A {\bf 366}, 27 (1981).
\bibitem{tsushima}
D. H. Lu, K. Tsushima, A. W. Thomas, A. G. Williams and K. Saito, 
Phys. Lett. B {\bf 441}, 27 (1998).
\bibitem{magdensity}
P. Roy Chowdhury, C. Samanta and D. N. Basu, Phys. Rev. C {\bf 73}, 014612 (2006); 
D. K. Srivastava, D. N. Basu and N. K. Ganguly, Phys. Lett. B {\bf 124}, 6 (1983).
\bibitem{WShypergm}
S. Fl\"ugge, {\it Practical Quantum Mechanics}, Springer (1998);
M. Ghominejad, Eur. Phys. J. Plus {\bf 128}, 59 (2013).
\bibitem{yorita}
T. Yorita {\it et al}., Phys. Lett. B {\bf 476}, 226 (2000).
\bibitem{bassthomas}
Steven D. Bass and Anthony W. Thomas, Phys. Lett. B {\bf 634}, 368 (2006). 
\bibitem{nogga}
A. Nogga, H. Kamada, W. Gl\"ockle and B. R. Barrett, Phys. Rev. C {\bf 65}, 054003 
(2002); 
A. Nogga, ``Nuclear and hypernuclear 3 and 4 body bound states", 
Ph. D. thesis, Ruhr Universit\"at, Bochum, 2001 (available at 
http://www-brs.ub.ruhr-uni-bochum.de/netahtml/HSS/
Diss/NoggaAndreas/).  
\end{thebibliography}
\end{document}